\documentclass[11pt,onecolumn,aps,showpacs,superscriptaddress,floatfix,
               preprintnumbers,amsmath,amssymb,nofootinbib]{revtex4}
\usepackage{graphicx}% Include figure files
\usepackage{dcolumn}% Align table columns on decimal point
\usepackage{bm}% bold math

\newcommand{\Z}{{\mathbb Z}}

\begin{document}
\title{Yukawa Corrections from Four-Point Functions\\ in Intersecting D6-Brane Models}

\author{Ching-Ming Chen}
\affiliation{George P. and Cynthia W. Mitchell Institute for
Fundamental Physics, Texas A\&M University,\\ College Station, TX
77843, USA}
\author{Tianjun Li}
\affiliation{George P. and Cynthia W. Mitchell Institute for
Fundamental Physics, Texas A\&M University,\\ College Station, TX
77843, USA} \affiliation{Institute of Theoretical Physics, Chinese
Academy of Sciences, Beijing 100080, China}
\author{V.~E. Mayes}
\affiliation{George P. and Cynthia W. Mitchell Institute for
Fundamental Physics, Texas A\&M University,\\ College Station, TX
77843, USA}
\author{D.~V. Nanopoulos}
\affiliation{George P. and Cynthia W. Mitchell Institute for
Fundamental Physics, Texas A\&M University,\\ College Station, TX
77843, USA} \affiliation{Astroparticle Physics Group, Houston
Advanced Research Center (HARC), Mitchell Campus,
Woodlands, TX~77381, USA; \\
Academy of Athens, Division of Natural Sciences, 28~Panepistimiou
Avenue, Athens 10679, Greece}

\preprint{ACT-03-08, MIFP-08-19}

\begin{abstract}

\begin{center}
\bf{Abstract}
\end{center}

We discuss corrections to the Yukawa  matrices of the Standard Model
(SM) fermions in intersecting D-brane models due to four-point
interactions. Recently, an intersecting D-brane model has been found
where it is possible to obtain correct masses and mixings for all
quarks as well as the tau lepton.  However, the masses for the first
two charged leptons come close to the right values but are not quite
correct. Since the electron and muon are quite light, it is likely
that there are additional corrections to their masses which cannot
be neglected. With this in mind, we consider contributions to the SM
fermion mass matrices from four-point interactions. In an explicit
model, we show that it is indeed possible to obtain the SM fermion
masses and mixings which are a better match to those resulting from
experimental data extrapolated at the unification scale when these
corrections are included. These corrections may have broader
application to other models.

\end{abstract}


\maketitle
\newpage

\section{Introduction}

In recent years, intersecting D-brane models, where the chiral
fermions arise at the intersections between D6-branes (Type IIA)
in the internal space~\cite{bdl} with the T-dual Type IIB
description in terms of magnetized D-branes~\cite{bachas} have
provided an exciting approach towards constructing semi-realistic
string vacua (for reviews, see~\cite{Blumenhagen:2005mu,
Blumenhagen:2006ci}). Indeed, such models provide promising setups
which may accommodate semi-realistic features of low-energy
physics. Given this, it is an interesting question to see how far
one can get from a particular string compactification to
reproducing the finer details of the Standard Model (SM) as a
low-energy effective field theory.

The Standard Model has an intricate structure, with
three-generations of chiral fermions which transform as
bifundamental representations of $SU(3)_C \times SU(2)_L \times
U(1)_Y$. In addition to the fact that the SM fermions are
replicated into three distinct generations, the different
generations exhibit a distinct pattern of mass hierarchies and
mixings. Interestingly, intersecting D-brane models may naturally
generate the SM fermion mass hierarchies and mixings, as well as
an explanation for the replication of chirality. In short,
D6-branes (in Type IIA) fill four-dimensional Minkowski space-time
and wrap 3-cycles in the compact manifold, with a stack of $N$
D6-branes having a gauge group $U(N)$ (or $U(N/2)$ in the case of
$\mathbf{T^6/(\Z_2 \times \Z_2)}$) in its world volume. The
3-cycles wrapped by the D-branes will in general intersect
multiple times in the internal space, resulting in chiral fermions
in the bifundamental representation localized at the intersections
between different stacks. The multiplicity of such fermions is
then given by the number of times the 3-cycles intersect.

The Yukawa couplings in intersecting D6-brane models arise from
open string world-sheet instantons that connect three D6-brane
intersections \cite{Aldazabal:2000cn}. For a given triplet of
intersections, the minimal world-sheet action which contributes to
the trilinear Yukawa couplings is weighted by a factor
$\exp(-A_{abc})$, where $A_{abc}$ is the world-sheet area of the
triangle bounded by the branes $a, b,$ and $c$. Since there are
several possible triangles with different areas, mass hierarchies
may inherently arise. The Yukawa couplings depend on both the
D-brane positions in the internal space as well as on the geometry
of the underlying compact manifold. Effectively, these quantities
are parameterized by the vacuum expectation values (VEVs) of open
and closed-string moduli.

Despite substantial progress in constructing semi-realistic vacua
with intersecting D-branes, there are many phenomenological
challenges remaining, besides the usual moduli stabilization
problem.   Typically, some or all of the Yukawa couplings are
typically forbidden by selection rules which arise from global
$U(1)$s which become massive via a generalized Green-Schwarz
mechanism.  It has been found that only Pati-Salam models can have
all the SM fermion Yukawa couplings present at the stringy tree
level, although some couplings which are perturbatively forbidden
may be generated in principle via D-brane instanton
effects~\cite{Ibanez:2006da, Blumenhagen:2006xt, Florea:2006si,
Abel:2006yk, Billo:2007py}. Also, there has generally been a rank
one problem in the SM fermion Yukawa matrices, preventing the
generation of masses and mixings for the first two families of
quarks and leptons. For the case of toroidal orientifold
compactifications, this can be traced to the fact that not all of
the SM fermions are localized at intersections on the same
torus~\cite{Cremades:2003qj, Chamoun:2003pf, Kitazawa:2004nf,
Dutta:2005bb}. However, one example of an intersecting D6-brane
model in Type IIA on the $\mathbf{T^6/(\Z_2\times \Z_2)}$
orientifold has recently been discovered in which these problems
may be solved~\cite{Cvetic:2004ui,Chen:2006gd}.  Thus, this
particular model may be a step forward to obtaining realistic
phenomenology from string theory. Indeed, as we have recently
shown~\cite{Chen:2007px}, it is possible within the moduli space
of this model to obtain the correct SM quark masses and mixings,
the tau lepton mass, and to generate naturally small neutrino
masses via the seesaw mechanism. In addition to these features,
the model exhibits automatic gauge coupling unification, and it is
possible to generate realistic low-energy supersymmetric particle
spectra, a subset of which may produce the observed dark matter
density.

In spite of the successes of this model, the electron and muon
masses come close to the right values, but they are still not
quite correct. Since the electron and muon masses are very light,
it is likely that there are additional corrections that must be
considered, as first suggested a long time ago in
\cite{Ellis:Nano}.  This idea was applied later in string theory,
for the case of the free-fermionic-formulation, where the general
rules for calculating higher order corrections, from
multipoint-functions, to the Yukawa couplings were first given in
\cite{Kalara:1990fb}. With this in mind, in this paper we consider
contributions to the Yukawa couplings from four-point functions.
Four-point string interactions in intersecting D6-brane scenarios
can be regarded as the scattering amplitudes of four string matter
fields \cite{Lust:2004cx}, and are consistent to the K\"ahler
metrics of twisted matter \cite{Bertolini:2005qh}. Both the
quantum and classical contributions have been studied in two
special cases where one has only an independent angle and the
other has two independent angles in a closed quadrilateral
\cite{Cvetic:Abel}. This was then promoted to the generalized
calculation of $N$-point amplitudes and four-point functions
without angle constraints \cite{Abel:2003yx}. This analysis is
based on the discussion of the twisted closed string interactions
on orbifolds \cite{Hamidi:1986vh, Dixon:1986qv, Burwick:1990tu},
where the analogy between open strings at brane intersections and
closed strings on orbifolds is known. The complete amplitudes for
Yukawa couplings, as a three-point limit of the four-point
amplitude from conformal field theory, are also discussed in
\cite{Lust:2004cx, Abel:2003yx, Cvetic:Abel}.

It is natural to ask whether the four-point function corrections
to the Yukawa couplings can generate the correct electron and muon
masses in the model discussed in \cite{Chen:2007px}; indeed the
correct electron and muon masses can be obtained by including
four-point function corrections in principle. However, the
D6-branes in this model can not form a closed quadrilateral for
the required four-point couplings on the relevant two torus. Two
stacks of D-branes overlap so the ``angle'' at their
``intersection'' turns out as a straight line. Instead of using
the conventional calculational procedure in this model, we can
still apply the idea of four-point function corrections by taking
the vector-like fields at the ``intersection'' on the overlapped
D6-branes in a asymptotic limit of the ``angle'' to $\pi$, a
physical mechanism which has not yet been clarified. Therefore, we
shall consider another model where the standard four-point
function corrections can be calculated. As we shall see later in
this model, if the SM fermion mass matrices receive contributions
from only three-point functions, one can only obtain the correct
quark masses and tau lepton mass, but it is not possible to
explain the CKM quark mixings, and the electron and muon masses.
Introducing corrections from four-point functions, one can indeed
obtain the correct quark and the tau lepton masses, and the CKM
quark  mixings, and as well as the electron mass. However, the
muon mass is still about 36\% smaller than the desired value.

In this paper, we review the construction of four-point functions,
and include both trilinear Yukawa couplings and four-point
interactions to obtain the SM fermion mass matrices in an explicit
intersecting D-brane model.  We show that it is indeed possible to
better match the masses and mixings for all fermions when four-point
corrections are included.  In addition, these corrections may have
broader application to other models.

\section{General Four-Point Functions}

Let us begin by considering an open string stretched between two
D-branes intersecting at an angle $\pi\theta$. From the boundary
conditions we can write the mode expansion as
\begin{eqnarray}
&&\partial X(z) = \sum_k \alpha_{k-\theta} z^{-k+\theta-1},
\nonumber \\
&&\partial \bar{X}(z) = \sum_k \bar{\alpha}_{k-\theta}
z^{-k-\theta-1},  \label{expansX}
\end{eqnarray}
where $z$ is the worldsheet coordinate.  By comparison of this
expression with the mode expansion for a closed string in the CFT
analysis of a $\Z_N$ orbifold twist field \cite{Hamidi:1986vh,
Dixon:1986qv, Burwick:1990tu}, the OPE can be written
as~\cite{Abel:2003yx}
\begin{eqnarray}
&&\partial X(z) \sigma_{\theta} (w,\bar{w}) \sim
(z-w)^{-(1-\theta)} \tau_{\theta} (w,\bar{w}), \nonumber \\
&&\partial \bar{X}(z) \sigma_{\theta} (w,\bar{w}) \sim
(z-w)^{-\theta} \tau'_{\theta} (w,\bar{w}),
\end{eqnarray}
where $X(w,\bar w)$ is the intersection point of the two
D-branes, $\sigma_{\theta} (w,\bar{w})$ is the open string twist field,
and $\tau$ and $\tau'$ are excited twist fields.  The local
monodromy conditions are then given by \cite{Abel:2003yx}
\begin{eqnarray}
&&\partial X(e^{2\pi i}(z-w)) = e^{2\pi i} \partial X(z-w), \nonumber \\
&&\partial \bar{X}(e^{2\pi i}(z-w)) = e^{-2\pi i} \partial
\bar{X}(z-w).
\end{eqnarray}

Since we are interested in a four-point interaction, we demand
that the four strings are in a bounded area such that $\sum
\theta_i =2$, and it is required that the four twist operators
$\sigma_{\theta_i}(z_i,\bar z_i)$ are present at the D-brane
intersections. Due to invariance under $SL(2,\mathbb{R})$, we can
set $z_1=0$, $z_2=x$, $z_3=1$, and $z_4=x_{\infty}$. The field $X$
has a classical piece $X_{cl}$ and a quantum piece $X_{qu}$, so
the interaction amplitude can be factorized into a classical
solution resulting from worldsheet instantons and a quantum
contribution resulting from quantum fluctuations as
\begin{equation}
Z= \sum_{\langle X_{cl}\rangle} e^{-S_{cl}} Z_{qu},
\end{equation}
where
\begin{equation}
S_{cl}=\frac{1}{4\pi \alpha'} \int d^2 z (\partial X_{cl}
\bar{\partial} \bar{X}_{cl} + \bar{\partial} X_{cl} \partial
\bar{X}_{cl}).
\end{equation}

\subsection{The Classical Contribution}

The classical field can be expanded by OPE as
\begin{eqnarray}
&&\partial X_{cl}(z) = a~ \omega(z), ~~ \partial \bar{X}_{cl}(z)
= \bar a~ \omega'(z), \nonumber \\
&&\bar{\partial} X_{cl}(\bar z) = b~ \bar \omega' (\bar z), ~~
\bar \partial \bar{X}_{cl}(\bar z) = \bar b~ \bar \omega (\bar z),
\end{eqnarray}
where
\begin{equation}
\omega(z)= \prod_i (z-x_i)^{-(1-\theta_i)}, ~~ \omega'(z)= \prod_i
(z-x_i)^{-\theta_i},  \label{omega}
\end{equation}
$a$, $\bar a$, $b$, and $\bar b$ are complex constants. On a
torus with brane intersection considered \cite{Abel:2003yx}, the
local monodromy condition is given by
\begin{equation}
X(e^{2\pi i} z, e^{-2\pi i}\bar z) = e^{2\pi i\theta}X +
(1-e^{2\pi i\theta})(f+v),
\end{equation}
where $f$ is the intersection point and $v$ is the lattice
translation of the torus.  Then from the global monodromy
conditions we have \cite{Abel:2003yx}
\begin{eqnarray}
\Delta_{C_i} X_{cl} = 4e^{-\pi i(\theta_i-\theta_{i+1})}
\sin\pi\theta_i \sin\pi\theta_{i+1} (f_{i+1}-f_i+v_i)= \oint_{C_i}
\big(dz \partial X_{cl}(z) + d\bar z \bar \partial X_{cl}(\bar
z)\big).
\end{eqnarray}

\begin{figure}[t]
\begin{center}
\includegraphics[width=.7\textwidth,angle=0]{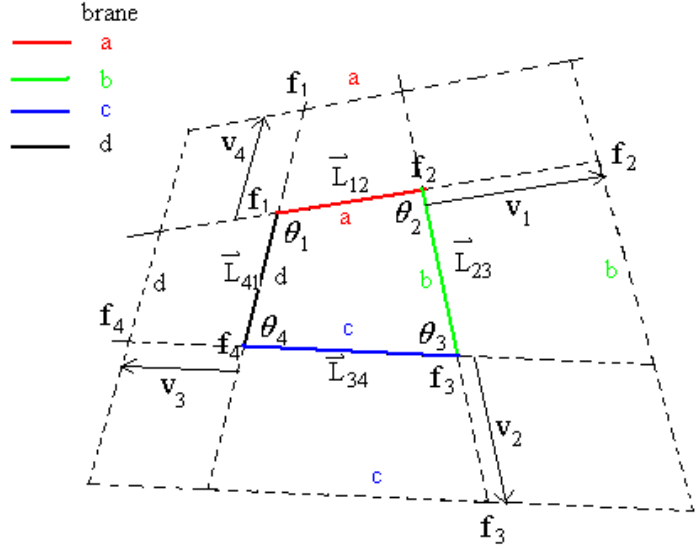}
\caption{Definition of the indices of the vertex and D-branes.}
\label{indexarea}
\end{center}
\end{figure}

In Fig. \ref{indexarea}, the index $i$ denotes the $i$-th vertex,
$f_1$ is the intersection point between stack $a$ and $b$ D-branes
and so on, and we define $\vec{L}_{i,i+1}$ to be the vector along
the direction from $f_i$ to $f_{i+1}$.  Thus $v_1=q_1|I_{ab}|
\vec{L}_{12}$, etc, where $q_i$ is an integer and $I_{ab}$ is the
intersection number. From the closure of the quadrilateral $\sum
v_i=0$, we can count the areas from the solutions of the linear
diophantine equations \cite{Abel:2003yx}
\begin{equation}
\left(\begin{array}{c}
q_1 I_{ba}  \\
q_2 I_{ba}  \\
\end{array} \right)= \left(\begin{array}{cc}
I_{cb} & I_{db}  \\
I_{ac} & I_{ad} \\
\end{array} \right) \left(\begin{array}{c}
q_3  \\
q_4 \\
\end{array} \right).
\end{equation}

In the case having two independent angles which we will consider
later for a specific model, if stacks $a$ and $c$ are parallel to
each other, then the above diophantine equations can be
simplified, and the four integer parameters $q_i$ are not
independent.  From \cite{Abel:2003yx} these $q_i$ can be
parameterized by two independent integer variables, which we will
use in the following area analysis by applying a different
approach.

Refer to \cite{Abel:2003yx} for the details of the calculation
involving hypergeometric functions from the integral of $\omega$
and $\omega'$, we can write down the minimum classical action when
there is non-zero worldsheet area on one subtorus as
\cite{Abel:2003yx}
\begin{equation}
S^{T^2}_{cl~min}= \frac{1}{2\pi\alpha'}\Bigg(\frac{\sin\pi\theta_1
\sin\pi\theta_4}{\sin(\pi\theta_1+\pi\theta_4)}\frac{v_{14}^2}{2}
+ \frac{\sin\pi\theta_2
\sin\pi\theta_3}{\sin(\pi\theta_2+\pi\theta_3)}\frac{v_{23}^2}{2}
\Bigg),
\end{equation}
where the equation in the parenthesis is exactly the formula for
the area of a quadrilateral with angles $\theta_i$ and two sides
$v_{14}$ and $v_{23}$.  We will then use this information to
calculate the classical contribution of the four-point functions.

\subsection{The Quantum Contribution}

The quantum contribution of the four-point function is given by
the correlator
\begin{equation}
\langle \sigma_{\theta_1}(z_1) \sigma_{\theta_2}(z_2)
\sigma_{\theta_3}(z_3) \sigma_{\theta_4}(z_4) \rangle.
\end{equation}
From the OPE of the stress tensor $T(z)$ with the twist fields we
have
\begin{equation}
T(z) \sigma_{\theta_j}(z_i)\sim \frac{h_j}{(z-z_i)^2} +
\frac{\partial_{z_i} \sigma_{\theta_j}(z_i)}{(z-z_i)} + \cdots,
\end{equation}
where $h_j=\frac{1}{2}\theta_j (1-\theta_j)$ is the conformal
dimension.  $T(z)$ also has a relation to $X$ as follows
\begin{equation}
-\frac{1}{2} \partial_z X \partial_w \bar{X} \sim
\frac{1}{(z-w)^2} + T(z) + \cdots.  \label{Tz-kinetic}
\end{equation}
Therefore, we can write
\begin{eqnarray}
&&\frac{\langle T(z) \sigma_{\theta_1}(z_1) \sigma_{\theta_2}(z_2)
\sigma_{\theta_3}(z_3) \sigma_{\theta_4}(z_4) \rangle }{\langle
\sigma_{\theta_1}(z_1) \sigma_{\theta_2}(z_2)
\sigma_{\theta_3}(z_3) \sigma_{\theta_4}(z_4) \rangle }
\nonumber\\
&=& \lim_{z \rightarrow w} \bigg[ \frac{\langle -\frac{1}{2}
\partial_z X \partial_w \bar{X}  \sigma_{\theta_1}(z_1)
\sigma_{\theta_2}(z_2) \sigma_{\theta_3}(z_3)
\sigma_{\theta_4}(z_4) \rangle }{\langle \sigma_{\theta_1}(z_1)
\sigma_{\theta_2}(z_2) \sigma_{\theta_3}(z_3)
\sigma_{\theta_4}(z_4) \rangle } - \frac{1}{(z-w)^2} \bigg].
\end{eqnarray}
We may then define the Green's function as
\begin{equation}
g(z,w;z_i) = \frac{\langle -\frac{1}{2}
\partial_z X \partial_w \bar{X}  \sigma_{\theta_1}(z_1)
\sigma_{\theta_2}(z_2) \sigma_{\theta_3}(z_3) \sigma_{\theta_4}(z_4)
\rangle }{\langle \sigma_{\theta_1}(z_1) \sigma_{\theta_2}(z_2)
\sigma_{\theta_3}(z_3) \sigma_{\theta_4}(z_4) \rangle }~.~\,
\end{equation}
From Eq. (\ref{Tz-kinetic}) we can find the asymptotic properties of
$g(z,w;z_i)$:
\begin{eqnarray}
g(z,w;z_i) &\sim& \frac{1}{(z-w)^2} + {\rm finite~~~~~~}{\rm for}~~~
z\rightarrow w, \nonumber \\
&\sim& \frac{1}{(z-z_i)^{-\theta_i}} ~~~~~~~~~~~  {\rm for}~~~z\rightarrow z_i, \nonumber \\
&\sim& \frac{1}{(w-z_i)^{-(1-\theta_i)}} ~~~~~~  {\rm
for}~~~w\rightarrow z_i.
\end{eqnarray}
Similarly from Eqs. (\ref{expansX}) and (\ref{omega}) we can write
down the equations for $X$ and $\bar X$:
\begin{eqnarray}
&&\partial X(z) \sim \omega_{\theta_i}(z) =
\prod(z-x_i)^{-(1-\theta_i)}, \nonumber \\
&&\partial \bar{X}(z) \sim \omega'_{\theta_i}(z) = \prod(z-x_i)^{
-\theta_i} =\omega_{1-\theta_i}(z).
\end{eqnarray}
Then $g(z,w;z_i)$ can be expanded in the following form
\cite{Abel:2003yx, Cvetic:Abel}
\begin{equation}
g(z,w;z_i) = \omega_{\theta_i}(z) \omega'_{\theta_i}(z) \bigg\{
\sum_{ij} a_{ij} \frac{(z-z_i)(z-z_j) \prod_k (w-z_k)
}{(w-z_i)(w-z_j)(z-w)^2} + A \bigg\}.
\end{equation}
The coefficients $a_{ij}$ can be fixed by the above asymptotic
relations, so finally we can determine
\begin{eqnarray}
&&\frac{\langle T(z) \sigma_{\theta_1}(z_1) \sigma_{\theta_2}(z_2)
\sigma_{\theta_3}(z_3) \sigma_{\theta_4}(z_4) \rangle }{\langle
\sigma_{\theta_1}(z_1) \sigma_{\theta_2}(z_2)
\sigma_{\theta_3}(z_3) \sigma_{\theta_4}(z_4) \rangle }
\nonumber\\
&=& -\frac{1}{2} \sum \theta_i\theta_j\frac{1}{(z-z_i)(z-z_j)}
+\frac{1}{2} \sum_{i<j} a_{ij}
(\frac{1}{z-z_i}+\frac{1}{z-z_j})^2+\frac{A}{\prod (z-z_i)}.
\end{eqnarray}

The quantum part of the monodromy conditions is independent of the
contours, so that
\begin{equation}
\Delta_{C_l} X_{qu} = 0 = \oint_{C_l} dz \partial X_{qu} +
\oint_{C_l} d\bar z \bar\partial X_{qu}. \label{monodromy}
\end{equation}
We can use this condition to determine $A$ and define two homology
cycles $C_1$ and $C_2$ by cutting the complex plane between $z_1$
and $z_2$, and between $z_3$ and $z_4$. In this way, we find
\begin{equation}
\oint_{C_i} dz g(z,w) + \oint_{C_i} d\bar{z} h(\bar{z},w) = 0,
\end{equation}
where $h(\bar{z},w)$ is the auxiliary correlation function
\begin{equation}
h(\bar{z},w;z_i)\equiv \frac{\langle -\frac{1}{2}
\partial_ {\bar{z}} X \partial_w \bar{X}  \sigma_{\theta_1}(z_1)
\sigma_{\theta_2}(z_2) \sigma_{\theta_3}(z_3)
\sigma_{\theta_4}(z_4) \rangle }{\langle \sigma_{\theta_1}(z_1)
\sigma_{\theta_2}(z_2) \sigma_{\theta_3}(z_3)
\sigma_{\theta_4}(z_4) \rangle }
=B\bar{\omega}_{1-\theta_i}(\bar{z}) \omega'_{\theta_i}(z).
\end{equation}
After using  the $SL(2,\mathbb{R})$ invariance, we obtain that  Eq.
(\ref{monodromy})  turns out to be
\begin{equation}
B \oint_{C_i} \bar\omega' (\bar z)d\bar z + A \oint_{C_i}
\omega(z) dz = x_{\infty} \oint_{C_i} \sum_i a_{i4} (z-x_i)
\omega(z) dz.
\end{equation}
Solving $A$, the correlator $\langle \sigma_{\theta_1}
\sigma_{\theta_2} \sigma_{\theta_3} \sigma_{\theta_4} \rangle$
then has a form \cite{Abel:2003yx}
\begin{equation}
\langle \sigma_{\theta_1} \sigma_{\theta_2} \sigma_{\theta_3}
\sigma_{\theta_4} \rangle = |I(x)|^{-\frac{1}{2}}
x_{\infty}^{-\theta_4 (1-\theta_4)} x^{\frac{1}{2}(\theta_1 +
\theta_2 -1)-\theta_1\theta_2} (1-x)^{\frac{1}{2}(\theta_2 +
\theta_3 -1)-\theta_2\theta_3},
\end{equation}
where $I(x)$ is a function of $x$, $\theta_i$, gamma functions of
$\theta_i$, and hypergeometric functions \cite{Abel:2003yx}.

Again, we are more interested in the case with two independent
angles, therefore by setting $\theta_1=1-\theta_2=\nu$ and
$\theta_4=1-\theta_3=\lambda$, the quantum part of the four-point
function can be written as \cite{Abel:2003yx, Cvetic:Abel}
\begin{equation}
Z_{4q} = 16 \pi^{\frac{5}{2}} x^{-\nu(1-\nu)} (1-x)^{-\nu\lambda}
I(x)^{-\frac{1}{2}}, \label{Z4qu}
\end{equation}
where
\begin{eqnarray}
I(x)=(1-x)^{(1-\nu-\lambda)} \big[B(\nu,\lambda)~_2F_1(\nu,
\lambda, \nu+\lambda; 1-x)~_2F_1(1-\nu, 1-\lambda, 1; x) \nonumber
\\+ B(1-\nu,1-\lambda)~_2F_1(1-\nu, 1-\lambda, 2-\nu-\lambda; 1-x)
~_2F_1(\nu, \lambda, 1; x)\big],
\end{eqnarray}
and $_2F_1$ is the hypergeometric function; $B(\nu,\lambda)=
\Gamma(\nu) \Gamma(\lambda)/\Gamma(\nu+\lambda)$. We have used the
fact that $_2F_1(\nu, \lambda, 1;
x)=(1-x)^{(1-\nu-\lambda)}~_2F_1(1-\nu, 1-\lambda, 1; x)$.  Note
that the quantum contribution Eq. (\ref{Z4qu}) remains invariant
under the interchange of the interior and exterior angles
$\nu,\lambda \leftrightarrow 1-\nu, 1-\lambda$.

The general form for a four-point amplitude on $\mathbf{T}^6$ with
two independent angles is then given~by
\begin{equation}
Z_4  = 16\pi^{5/2} \prod_j x_j^{-\nu_j(1-\nu_j)}
(1-x_j)^{-\nu_j\lambda_j} I_j(x_j)^{-1/2} \sum_{k,l} {\exp}
\big(-\frac{A_{4j}}{2\pi\alpha'}\big) . \label{Zq4}
\end{equation}

By taking the limit of $x$, $Z_{3q}$ is fixed to a constant and we
obtain the three-point amplitude for the Yukawa couplings
\cite{Cvetic:Abel, Lust:2004cx}
\begin{equation}
Z_3  = 2\pi \prod_j [\frac{16\pi^2 \Gamma(1-\nu_j)
\Gamma(1-\lambda_j) \Gamma(\nu_j+\lambda_j)}{\Gamma(\nu_j)
\Gamma(\lambda_j) \Gamma(1-\nu_j-\lambda_j)}]^{\frac{1}{4}} \sum_m
{\exp} \big(-\frac{A_j}{2\pi\alpha'}\big).  \label{Zq3}
\end{equation}

As we will later encounter in the model considered in the next
section, where only the D-branes on the second torus form closed
areas either for four-point or three-point amplitudes, the quantum
contributions from other tori are just constants, which are then
able to be absorbed into the VEVs of the Higgs fields.  Using a
computer code,  we find that the ratio of $Z_{4q}$ to $Z_{3q}$ is
around $\mathcal{O}(10)$ which can then also be absorbed into the
VEVs for simplicity.  Therefore, we can merely focus on the
classical contributions to the SM fermion masses and mixings.

\section{A Working Example of the SM Fermion Masses and Mixings}

\subsection{The Pati-Salam Models}

Let us first review the Pati-Salam model discussed in
\cite{Chen:2007px}, where its D6-brane configurations and
intersecting numbers are presented in Table~\ref{real}.  To
explain the electron and muon masses in this model, we are looking
for four-point interactions such as
\begin{equation}
\phi^i_{ab} \phi^j_{ca} \phi^k_{b'c} \phi^l_{bb'} ~~\mathrm{or}~~
\phi^i_{ab} \phi^j_{ca} \phi^k_{cc'} \phi^l_{bc'}~,~\,
\end{equation}
where $\phi^i_{\alpha \beta}$ are the chiral superfields at the
intersections between stack $\alpha$ and $\beta$ D6-branes.

\begin{table}[h]
\begin{center}
\footnotesize
\begin{tabular}{|@{}c@{}|c||@{}c@{}c@{}c@{}|} \hline

stk & $N$ & ($n_1$, $l_1$) & ($n_2$, $l_2$) & ($n_3$, $l_3$)    \\
\hline \hline

$a$ & 8 & ( 0,-1) & ( 1, 1) & ( 1, 1)  \\ \hline

$b$ & 4 & ( 3, 1) & ( 1, 0) & ( 1,-1)    \\ \hline

$c$ & 4 & ( 3,-1) & ( 0, 1) & ( 1,-1)
\\ \hline  %\hline
\end{tabular}
\caption{D6-brane configurations and intersecting numbers for the
model presented in~\cite{Chen:2007px}, where the SM fermions and
Higgs fields are from the intersections on the first torus. This
model is constructed from Type IIA ${\mathbf T^6/(\Z_2\times
\Z_2)}$ orientifold.} \label{real}
\end{center}    %T6-1-(U3)
\end{table}

From Table~\ref{real}, considering the wrapping numbers on the
first two-torus, we find that the $\Omega R$ image $b'$ of stack
$b$ is parallel to stack $c$, and stack $b$ is parallel to the
$\Omega R$ image $c'$ of stack $c$. It is required that $b$ and
$c'$ or $c$ and $b'$ must overlap to form a closed area, which is
not exactly forming a quadrilateral.  After carefully studying the
possible four-point functions on the first torus, we find that we
can obtain the correct electron and muon masses. However, the
techniques for calculating such kind of geometric structure for
four-point functions have not been clarified yet. Therefore, we
turn our attention to the other models. To avoid the generic rank
one problem, we require that the three SM fermion families arise
from the intersections on the same two-torus.  The number of the
Higgs bidoublets from that torus is a multiple of 3. In addition,
we require that the $\Omega R$ image of the $U(2)_L$ D-brane stack
forms a closed quadrilateral with the $U(4)$, $U(2)_L$, and
$U(2)_R$ stacks to get rid of exotic particles. And there may also
arise additional constraints on the D-brane wrapping numbers to
form a bounded area. Generally a three-generation model without
exotic SM type particles has a structure like that of a stack of
D-branes which lies on the orbifold while the other two are images
of each other and the ratio of the wrapping numbers is three, for
example $(n,m)=(1,3)$, just as the model discussed in
\cite{Chen:2007px}. However, similar to the above model, we find
that such types of models cannot form a normal bounded four-point
area. Thus, we will consider the Model TI-U-3 in
\cite{Chen:2006gd}. We present its D6-brane configurations and
intersecting numbers of the observable sector in Table~\ref{T1},
and the particle spectrum of the observable sector in Table
\ref{Spectrum-I}, where $S_L^i$ and $\overline{S}_L^i$ are the SM
singlets from the intersections of stack $b$ and its $\Omega R$
image~$b'$.

\begin{table}[h]
\begin{center}
\footnotesize
\begin{tabular}{|@{}c@{}|c||@{}c@{}c@{}c@{}||c|c||c|@{}c@{}|@{}c@{}|@{}c@{}||
@{}c@{}|@{}c@{}|} \hline

stk & $N$ & ($n_1$, $l_1$) & ($n_2$, $l_2$) & ($n_3$, $l_3$) & A &
S & $b$ & $b'$ & $c$ & $c'$ & $d$ & $O6$    \\
\hline \hline

$a$ & 4 & ( 1, 1) & ( 1,-3) & ( 1, 0) & 0 & 0 & 3 & 0(1) & -3 &
0(3) & -3 & 0(3) \\ \hline

$b$ & 2 & ( 2, 0) & ( 1, 3) & ( 1,-1) & 0 & 0 & - & - & 6 & 0(3) &
6 & 0(3)   \\ \hline

$c$ & 2 & ( 1,-1) & ( 2, 0) & ( 1, 1) & 0 & 0 & - & - & - & - &
0(1) & 0(1)
\\ \hline  %\hline

%$d$ & 2 & ( 0,-2) & ( 2, 0) & ( 0, 1) & - & - & - & - & - & - & -
%& 0(2) \\ \hline

%$O6$ & 1 & ( 2, 0) & ( 2, 0) & ( 1, 0) & - & - &
%\multicolumn{6}{|c|}{$2\chi_3=3\chi_2=\chi_1$}
% \\ \hline
\end{tabular}
\caption{D6-brane configurations and intersecting numbers in the
Model TI-U-3 in \cite{Chen:2006gd}, where the SM fermions and
three of the six Higgs fields are from the intersections on the
second two-torus. This model is constructed in the supersymmetric
AdS vacuum on Type IIA ${\mathbf T^6}$ orientifold with flux
compactifications.} \label{T1}
\end{center}    %T6-1-(U3)
\end{table}

\begin{table}[htb]
\footnotesize
\renewcommand{\arraystretch}{1.0}
\begin{center}
\begin{tabular}{|c||c||c|c|c||c|c|c|}\hline
 & Quantum Number
& $Q_4$ & $Q_{2L}$ & $Q_{2R}$  & Field \\
\hline\hline
$ab$ & $3 \times (4,\bar{2},1,1)$ & 1 & -1 & 0  & $F_L(Q_L, L_L)$\\
$ac$ & $3\times (\bar{4},1,2,1)$ & -1 & 0 & $1$   & $F_R(Q_R, L_R)$\\
\hline\hline
$ac'$ & $3 \times (4,1,2,1)$ & 1 & 0 & 1  & $\Phi_i$ \\
& $3 \times (\bar{4}, 1, \bar{2},1)$ & -1 & 0 & -1 &
$\overline{\Phi}_i$\\
\hline
$bc$ & $6 \times (1,2,\overline{2},1)$ & 0 & 1 & -1   & $H_u^i$, $H_d^i$\\
& $6 \times (1,\overline{2},2,1)$ & 0 & -1 & 1   & \\
\hline
$b'c$ & $3 \times (1,\overline{2}, \overline{2},1)$ & 0 & -1 & -1   & $H'_u$, $H'_d$\\
& $3 \times (1,2, 2,1)$ & 0 & 1 & 1   & \\
\hline
$b'b$
 & $6 \times (1, 1, 1,1)$ & 0 & 2 & 0   & $S_L^i$ \\
 & $6 \times (1, \overline{1}, 1,1)$ & 0 & -2 & 0   & $\overline{S}_L^i$ \\
\hline
\end{tabular}
\caption{The chiral and vector-like superfields in the observable
sector, and their quantum numbers under the gauge symmetry
$SU(4)_C\times SU(2)_L\times SU(2)_R$.} \label{Spectrum-I}
\end{center}
\end{table}

The superpotential which includes the trilinear Yukawa couplings
is given by
\begin{eqnarray}
\mathcal{W}_3 \sim  Y^u_{ijk} Q^i  U^{cj} H^k_u + Y^d_{ijk} Q^i
D^{cj} H^k_d +  Y^l_{ijk} L^i  E^{cj} H^k_d~,~\,
\end{eqnarray}
where $Y^u_{ijk}$, $Y^d_{ijk}$, and $Y^l_{ijk}$ are Yukawa
couplings, and $Q^i$, $U^{ci}$, $D^{ci}$, $L^i$ and $E^{ci}$ are
the left-handed quark doublet, right-handed up-type quarks,
right-handed down-type quarks, left-handed lepton doublet, and
right-handed leptons, respectively. The superpotential including
the four-point interactions is
\begin{eqnarray}
\mathcal{W}_4 \sim {1\over {M_S}} \left( Y^{\prime u}_{ijkl}
 Q^i U^{cj} H^{\prime k}_u S_L^l +
Y^{\prime d}_{ijkl} Q^i D^{cj} H^{\prime k}_d S_L^l + Y^{\prime
l}_{ijkl} L^i E^{cj} H^{\prime k}_d S_L^l \right)~,~\,
\end{eqnarray}
where $Y^{\prime u}_{ijkl}$, $Y^{\prime d}_{ijkl}$, and $Y^{\prime
l}_{ijkl}$ are Yukawa couplings of the four-point functions, and
$M_S$ is the string scale.

\subsection{Yukawa Couplings from Three-Point Functions}

In order to calculate the trilinear Yukawa couplings, we first note
the intersection numbers on three two-tori
\begin{eqnarray}
&& I_{ab}^{(1)}=-1 ~,~~~ I_{ab}^{(2)}= 3 ~,~~~
I_{ab}^{(3)}=-1~,~\nonumber  \\
&& I_{ca}^{(1)}= 1 ~,~~~ I_{ca}^{(2)}=-3 ~,~~~ I_{ca}^{(3)}=-1~,~\nonumber  \\
&& I_{bc}^{(1)}=-1~,~~~ I_{bc}^{(2)}=-3 ~,~~~ I_{bc}^{(3)}= 2~.~\,
\label{intersection numbers}
\end{eqnarray}
From these, we find that
$d^{(2)}=g.c.d.(I_{ab}^{(2)},I_{bc}^{(2)},I_{ca}^{(2)})=3$,
$d^{(1)}=1$, and $d^{(3)}=1$. The parameters of the theta
functions in terms of the intersection numbers, the brane shifts
$\epsilon^{(i)}_{\alpha}$, the Wilson line phases
$\theta^{(i)}_{\alpha}$, and the K\"ahler moduli $J^{(i)}$ are
\cite{Cremades:2003qj}
\begin{eqnarray}
&&\delta^{(1)} = -\epsilon_c^{(1)} + \epsilon_b^{(1)} -
\epsilon_a^{(1)},  \nonumber  \\
&&\delta^{(2)} = \frac{i^{(2)}}{3} - \frac{j^{(2)}}{3} -
\frac{k^{(2)}}{3} + \frac{\epsilon_c^{(2)} - \epsilon_b^{(2)} -
\epsilon_a^{(2)}}{3} + \frac{s^{(2)}}{3}, \nonumber \\
&&\delta^{(3)} = \frac{k^{(3)}}{2} + \frac{-\epsilon_c^{(3)} -
\epsilon_b^{(3)} + 2\epsilon_a^{(3)}}{2} + \frac{s^{(3)}}{2},
\end{eqnarray}
\begin{eqnarray}
&&\phi^{(1)} = -\theta_c^{(1)} + \theta_b^{(1)}
- \theta_a^{(1)}, \nonumber \\
&&\phi^{(2)} =~~\theta_c^{(2)} - \theta_b^{(2)} - \theta_a^{(2)} ,
\nonumber \\
&&\phi^{(3)} =-\theta_c^{(3)} - \theta_b^{(3)}+2 \theta_a^{(2)},
\end{eqnarray}
\begin{eqnarray}
\kappa^{(1)} = \frac{J^{(1)}}{\alpha'}, ~~ \kappa^{(2)} =
\frac{3J^{(3)}}{\alpha'}, ~~ \kappa^{(3)} =
\frac{2J^{(3)}}{\alpha'}.
\end{eqnarray}

For convenience we redefine the shift on each torus as
\begin{eqnarray}
\epsilon^{(1)} \equiv -\epsilon_c^{(1)} + \epsilon_b^{(1)} -
\epsilon_a^{(1)}~,~~ \epsilon^{(2)} \equiv \frac{\epsilon_c^{(2)} -
\epsilon_b^{(2)} - \epsilon_a^{(2)}}{3}~,~~ \epsilon^{(3)} \equiv
\frac{-\epsilon_c^{(3)} - \epsilon_b^{(3)} +
2\epsilon_a^{(3)}}{2}~,~\,
 \label{shifts}
\end{eqnarray}
then the theta function of each torus can be written as
\cite{Cremades:2003qj}
\begin{equation}
\vartheta \left[\begin{array}{c} \delta^{(r)}\\ \phi^{(r)}
\end{array} \right] (\kappa^{(r)})=\sum_{l_r\in\mathbf{Z}} e^{\pi
i(\delta^{(r)}+l_r)^2 \kappa^{(r)}} e^{2\pi i(\delta^{(r)}+l_r)
\phi^{(r)}}  ~,~\, \label{Dtheta}
\end{equation}
where $r=1,2,3$, so the Yukawa coupling constant can be expressed
as
\begin{equation}
Y_{\{ijk\}}=Z_{3q} \sigma_{abc} \prod_{r=1}^3 \vartheta
\left[\begin{array}{c} \delta^{(r)}\\ \phi^{(r)}
\end{array} \right] (\kappa^{(r)})~,~\,
\end{equation}
where $Z_{3q}$ stands for the quantum contribution to the
instanton amplitude, and $\sigma_{abc}=\prod_r
\mathrm{sign}(I_{ab}^{(r)}I_{bc}^{(r)}I_{ca}^{(r)})$.

For the first torus the intersection number is one, so its
contribution is just a constant. On the second torus $i^{(2)},
j^{(2)}, k^{(2)}$ range from $0-2$, and on the third torus
$k^{(3)}$ goes from 0 to 1. Thus, although the total number of
Higgs state is $k^{(2)}\times k^{(3)}=3\times 2=6$, only three
linear combinations of six Higgs fields can provide the SM fermion
Yukawa couplings since the intersection number between $b$ and $c$
stacks of D6-brane is 3 on the second two-torus. Explicitly, there
is only one SM Higgs linear combination from the two bidoublet
fields $(k^{(2)}, k^{(3)}) = \{ (k^{(2)}, 0), (k^{(2)}, 1) \}$.
Furthermore, since the triplet of intersections is connected by an
instanton, the selection rule for the indices of the second torus
\begin{equation}
i^{(2)} + j^{(2)} + k^{(2)} = 0\; \mathrm{ mod }\;
3~,~\,
\label{select}
\end{equation}
should be satisfied.  Then we can choose the Yukawa coupling
matrices as the following form
\begin{eqnarray}
Y^{(2)}_{k=0}   \sim  \left(\begin{array}{ccc}
A & 0 & 0 \\
0 & 0 & C \\
0 & B & 0  \end{array} \right),                      \;\;
Y^{(2)}_{k=1} \sim  \left(\begin{array}{ccc}
0 & 0 & B \\
0 & A & 0 \\
C & 0 & 0  \end{array} \right),                      \;\;
Y^{(2)}_{k=2} \sim  \left(\begin{array}{ccc}
0 & C & 0 \\
B & 0 & 0 \\
0 & 0 & A  \end{array} \right),
\end{eqnarray}
where
\begin{eqnarray}
A \equiv \vartheta \left[\begin{array}{c} \epsilon^{(2)}\\
\phi^{(2)}   \end{array}  \right]  (\frac{3J^{(2)}}{\alpha'}), ~~
B \equiv \vartheta \left[\begin{array}{c} \epsilon^{(2)}+
\frac{1}{3}\\ \phi^{(2)} \end{array} \right]
(\frac{3J^{(2)}}{\alpha'}),~~ C \equiv \vartheta
\left[\begin{array}{c} \epsilon^{(2)}-\frac{1}{3}\\  \phi^{(2)}
\end{array} \right] (\frac{3J^{(2)}}{\alpha'}),
\end{eqnarray}
where we simply set here $s^{(2)}=k^{(2)}$. Similarly in the third
torus,  there is only one parameter
\begin{eqnarray}
Y^{(3)}_{k=0} \sim \vartheta \left[\begin{array}{c}
\epsilon^{(3)}\\ \phi^{(3)} \end{array} \right]
(\frac{2J^{(3)}}{\alpha'}) \equiv A_3, ~~ Y^{(3)}_{k=1} \sim
\vartheta \left[\begin{array}{c} \epsilon^{(3)}+\frac{1}{2}\\
\phi^{(3)}
\end{array} \right] (\frac{2J^{(3)}}{\alpha'}) \equiv D_3.
\end{eqnarray}

Therefore the classical part of this three-point couplings is given
by
\begin{eqnarray}
&&Z_{3cl1}  = \left(\begin{array}{@{}ccc@{}}
A A_3 & 0 & 0 \\
0 & 0 & C A_3 \\
0 & B A_3 & 0  \end{array} \right),                           \;\;
Z_{3cl2} = \left(\begin{array}{@{}ccc@{}}
0 & 0 & B A_3 \\
0 & A A_3 & 0 \\
C A_3 & 0 & 0  \end{array} \right),                           \;\;
Z_{3cl3} = \left(\begin{array}{@{}ccc@{}}
0 & C A_3 & 0 \\
B A_3 & 0 & 0 \\
0 & 0 & A A_3  \end{array} \right),  \nonumber \\
&&Z_{3cl4} = \left(\begin{array}{@{}ccc@{}}
A D_3 & 0 & 0 \\
0 & 0 & C D_3 \\
0 & B D_3 & 0  \end{array} \right),                           \;
Z_{3cl5} = \left(\begin{array}{@{}ccc@{}}
0 & 0 & B D_3 \\
0 & A D_3 & 0 \\
C D_3 & 0 & 0  \end{array} \right),                           \;
Z_{3cl6} = \left(\begin{array}{@{}ccc@{}}
0 & C D_3 & 0 \\
B D_3 & 0 & 0 \\
0 & 0 & A D_3  \end{array} \right)~.
\end{eqnarray}
If each of the Higgs states resulting from intersections localized
between stacks $b$ and $c$ develops a VEV, $v_i^{\Phi}$, where
$i=1,\cdots,6$  for each of the six Higgs states, and $\Phi$ is an
index for $(u)$p-type quarks, $(d)$own-type quarks, and
$(l)$eptons, respectively, the total effects of Yukawa couplings
including the quantum part will be
\begin{eqnarray}
Z_3^{\Phi}  = Z_{3q} \left(\begin{array}{@{}ccc@{}}
A (A_3v_1^{\Phi}+D_3v_4^{\Phi}) & C (A_3v_3^{\Phi}+D_3v_6^{\Phi}) & B (A_3v_2^{\Phi}+D_3v_5^{\Phi}) \\
B (A_3v_3^{\Phi}+D_3v_6^{\Phi}) & A (A_3v_2^{\Phi}+D_3v_5^{\Phi}) & C (A_3v_1^{\Phi}+D_3v_4^{\Phi}) \\
C (A_3v_2^{\Phi}+D_3v_5^{\Phi}) & B (A_3v_1^{\Phi}+D_3v_4^{\Phi})
& A (A_3v_3^{\Phi}+D_3v_6^{\Phi})  \end{array} \right)~,~\,
\label{Yukawa}
\end{eqnarray}
where $v_i^{(d)} = v_i^{(l)}$. Thus, it is clear that only three
linear combinations of the six Higgs states contribute to the
Yukawa couplings: $A_3H_1^{\Phi}+D_3H_4^{\Phi}$,
$A_3H_3^{\Phi}+D_3H_6^{\Phi}$, and $A_3 H_2^{\Phi}+D_3
H_5^{\Phi}$, where for simplicity we neglect the normalization.

\subsection{Yukawa Couplings from Four-Point Functions}

The formula for the area of a quadrilateral in terms of its angles
and two sides and the solutions of diophantine equations for
estimating the multiple areas of the quadrilaterals from non-unit
intersection numbers are given in~\cite{Abel:2003yx}. In the
present discussion, we are considering a model which possesses
only two independent angles, so we need at least two parameters to
describe all quadrilaterals. In addition to these formulae, there
is a more intuitive way to calculate the area for these four-sided
polygons with only two independent angles. A quadrilateral with
two independent angles is a trapezoid, and a trapezoid can be
always taken as the difference between two similar triangles.
Therefore, since we know the classical part is
\begin{equation}
Z_{4cl} \sim e^{-A_{quad}},
\end{equation}
it is equivalent to write
\begin{equation}
Z_{4cl} \sim e^{-|A_{tri}-A'_{tri}|}.
\end{equation}

\begin{figure}[h]
\begin{center}
\includegraphics[width=.5\textwidth,angle=0]{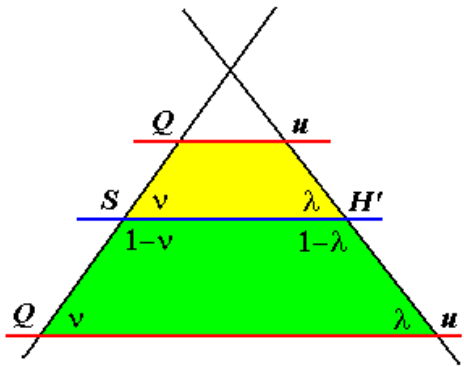}
\end{center}
\caption{A picture of two trapezoids with different field orders.
The red brane repeats in a next cycle and can still form a similar
trapezoid with the blue brane. This coupling also contributes to
the four-point function.} \label{2-cases}
\end{figure}

Taking the absolute value of the difference reveals that there are
two cases: $A_{tri}>A'_{tri}$ and $A_{tri}<A'_{tri}$, as shown in
Fig. \ref{2-cases}.  From the figure we can see the two trapezoids
are similar with different sizes, but the orders of the fields
corresponding to the angles are different, which is under an
interchange of $\theta \leftrightarrow 1-\theta$, $\theta=\nu,
\lambda$. These different field orders may cause different values
for their quantum contributions. However, we have shown above that
this angle transformation will not affect the quantum contribution,
so these two cases are on equal foot to sum up. Therefore, we are
able to employ the same techniques which have developed for
calculating the trilinear Yukawa couplings.

\begin{figure}[h]
\begin{center}
\includegraphics[width=.5\textwidth,angle=0]{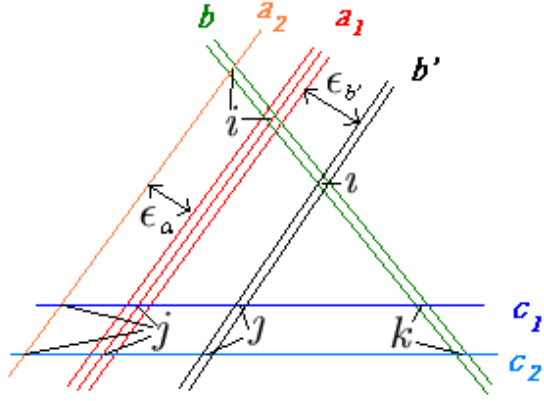}
\end{center}
\caption{A diagram showing the areas bounded by stacks of D-branes
which give rise to the Yukawa couplings for quarks and leptons via
world-sheet instantons.  The Yukawa couplings of the up-type
quarks are from the areas by stack $a_1$, $b$, $c_1$, the
down-type quarks by stack $a_1$, $b$, $c_2$, and the leptons by
$a_2$, $b$, $c_2$.  The four-point function corrections to the
Yukawa couplings of the up-type quarks are from the areas by stack
$a_1$, $b$, $b'$, $c_1$, the down-type quarks by stack $a_1$, $b$,
$b'$, $c_2$, and the leptons by $a_2$, $b$, $b'$, $c_2$. }
\label{brane area}
\end{figure}

For a trapezoid formed by the stacks $a$, $b$, $b'$, $c$, we can
calculate it as the difference between two triangles formed by
stacks $a$, $b$, $c$ and $b'$, $b$, $c$. In other words, they
share the same intersection $I_{bc}$.  Therefore, if we use this
method to calculate the trapezoidal area, we should keep in mind
that the intersection index $k$ for $I_{bc}$ remains the same for
a certain class of trapezoids when varying other intersecting
indices. Here we set indices $i$ for $I_{ab}$, $j$ for $I_{ca}$,
$\imath$ for $I_{b'b}$, and $\jmath$ for $I_{cb'}$, as shown in
Fig. \ref{brane area}. We may calculate the areas of the triangles
as we did in the trilinear Yukawa couplings
above~\cite{Cremades:2003qj}
\begin{eqnarray}
&&A_{ijk}=\frac{1}{2}(2\pi)^2 A_{\mathbf{T^2}}
|I_{ab}I_{bc}I_{ca}|~ \big(\frac{i}{I_{ab}} +\frac{j}{I_{ca}}
+\frac{k}{I_{bc}}
+\epsilon +l \big)^2, \nonumber \\
&&A_{\imath\jmath k}=\frac{1}{2}(2\pi)^2 A_{\mathbf{T^2}}
|I_{b'b}I_{bc}I_{cb'}|~ \big( \frac{\imath}{I_{b'b}}
+\frac{\jmath}{I_{cb'}} +\frac{k}{I_{bc}} +\varepsilon +\ell
\big)^2,  \label{Tareas}
\end{eqnarray}
where $i$, $j$, $k$ and $\imath$, $\jmath$, $k$ are using the same
selection rules as Eq. (\ref{select}). Thus, the classical
contribution of the four-point functions is given by
\begin{equation}
Z_{4cl} = \sum_{l,\ell} e^{-\frac{1}{2\pi}|A_{ijk}-A_{\imath\jmath
k}|}.
\end{equation}
Note that this formula will diverge when $A_{ijk}=A_{\imath\jmath
k}$, which is due to over-counting the zero area when the
corresponding parameters in Eq. (\ref{Tareas}) are the same.  In
such a case, $Z_{4cl} = 1+\sum_{l\neq\ell}
e^{-\frac{1}{2\pi}|A_{ijk}-A_{\imath\jmath k}|}$.  We will not
meet this special situation in our following discussion.

In the model of Table~\ref{T1}, in addition to the intersection
numbers in Eq.~(\ref{intersection numbers}), we have
\begin{eqnarray}
&& I_{bb'}^{(1)}= 0 ~,~~~ I_{bb'}^{(2)}=-3 ~,~~~
I_{bb'}^{(3)}=~2~;~\nonumber  \\
&&I_{bO6}^{(1)}= 0\,,~~~I_{bO6}^{(2)} = -3\,,~~~ I_{bO6}^{(3)} =
2~;~~~ \nonumber \\
&& I_{cb'}^{(1)}= 1 ~,~~~ I_{cb'}^{(2)}=-3 ~,~~~ I_{cb'}^{(3)}=0~,
\end{eqnarray}
and the number of the singlets from the $SU(2)_L$ anti-symmetric
representation is given by $\mathcal{M}_{Anti}=\prod_i
(I^{(i)}_{bb'}+I^{(i)}_{bO6})/2$.  It is obvious that we have six
vector-like anti-symmetric fields for $SU(2)_L$, three from the
second torus and two from the third torus.  The matrix elements on
the second torus from the four-point functions can be written in
terms of $a_{ij\imath\jmath}$ as
\begin{eqnarray}
&&\left(\begin{array}{ccc}
a_{0000} & 0 & 0 \\
0 & 0 & a_{1200} \\
0 & a_{2100} & 0  \end{array} \right),                      \;\;
\left(\begin{array}{ccc}
0 & a_{0101} & 0 \\
a_{1001} & 0 & 0 \\
0 & 0 & a_{2201}  \end{array} \right),                   \;\;
\left(\begin{array}{ccc}
0 & 0 & a_{0202} \\
0 & a_{1102} & 0 \\
a_{2002} & 0 & 0  \end{array} \right),       \nonumber \\
&&\left(\begin{array}{ccc}
0 & a_{0110} & 0 \\
a_{1010} & 0 & 0 \\
0 & 0 & a_{2210}  \end{array} \right),                      \;\;
\left(\begin{array}{ccc}
0 & 0 & a_{0211} \\
0 & a_{1111} & 0 \\
a_{2011} & 0 & 0  \end{array} \right),                      \;\;
\left(\begin{array}{ccc}
a_{0012} & 0 & 0 \\
0 & 0 & a_{1212} \\
0 & a_{2112} & 0  \end{array} \right),            \nonumber \\
&&\left(\begin{array}{ccc}
0 & 0 & a_{0220} \\
0 & a_{1120} & 0 \\
a_{2020} & 0 & 0  \end{array} \right),                      \;\;
\left(\begin{array}{ccc}
a_{0021} & 0 & 0 \\
0 & 0 & a_{1221} \\
0 & a_{2121} & 0  \end{array} \right),             \;\;
\left(\begin{array}{ccc}
0 & a_{0122} & 0 \\
a_{1022} & 0 & 0 \\
0 & 0 & a_{2222}  \end{array} \right).                      \;\;
\end{eqnarray}
The relations between these elements are shown in Table~\ref{4-area}.

\begin{table}[h]
\begin{center}
\footnotesize
\begin{tabular}{|c|c|c||c|c|c|} \hline

Parameters & $\delta$ & $d$ & $k=0$ & $k=1$ & $k=2$ \\
\hline

$\mathcal{A}$ & 0 & 0 & $a_{0000}$ & $a_{1111}$ & $a_{2222}$ \\
\hline

$\mathcal{B}$ & $\frac{1}{3}$ & 0 & $a_{2100}$ & $a_{0211}$ & $a_{1022}$ \\
\hline

$\mathcal{C}$ & $\frac{-1}{3}$ & 0 & $a_{1200}$ & $a_{2011}$ & $a_{0122}$ \\
\hline

$\mathcal{D}$ & 0 & $\frac{1}{3}$ & $a_{0021}$ & $a_{1102}$ & $a_{2210}$ \\
\hline

$\mathcal{E}$ & $\frac{1}{3}$ & $\frac{1}{3}$ & $a_{2121}$ & $a_{0202}$ & $a_{1010}$ \\
\hline

$\mathcal{F}$ & $\frac{-1}{3}$ & $\frac{1}{3}$ & $a_{1221}$ &
$a_{2002}$ & $a_{0110}$ \\ \hline

$\mathcal{G}$ & 0 & $\frac{-1}{3}$ & $a_{0012}$ & $a_{1120}$ &
$a_{2201}$ \\ \hline

$\mathcal{H}$ & $\frac{1}{3}$ & $\frac{-1}{3}$ & $a_{2112}$ &
$a_{0220}$ & $a_{1001}$ \\ \hline

$\mathcal{I}$ & $\frac{-1}{3}$ & $\frac{-1}{3}$ & $a_{1212}$ &
$a_{2020}$ & $a_{0101}$ \\ \hline

\end{tabular}
\caption{The matrix elements in terms of parameters.
$\delta=\frac{i}{I_{ab}} +\frac{j}{I_{ca}} +\frac{k}{I_{bc}}$ and
$d=\frac{\imath}{I_{b'b}} +\frac{\jmath}{I_{cb'}}
+\frac{k}{I_{bc}}$. Note that since we need to sum over $l$ and
$\ell$, an integer shift has no effect, so $\frac{1}{3}$ is
equivalent to $\frac{-2}{3}$; $\frac{-1}{3}$ is equivalent to
$\frac{2}{3}$ for $d$ and $\delta$, etc.  } \label{4-area}
\end{center}    %T6-1-(U3)
\end{table}
In this model, we have six SM singlet fields $S_L^i$ and three
Higgs-like states $H^{\prime i}_{u, d}$. Similar to the Higgs
fields $H^{i}_{u, d}$, only three linear combinations of the six
$S_L^i$ can contribute to the four-point Yukawa couplings. Thus,
the four-point Yukawa couplings involve three SM singlet fields
$S_L^i$ and three Higgs-like states $H^{\prime i}_{u, d}$. If
their VEVs are denoted as $u_{\imath}$ and $w_{\jmath}$
respectively, then the complete contribution will be
\begin{eqnarray}
Z_{4cl}        \sim \left(\begin{array}{@{}ccc@{}}
\mathcal{A}u_1w_1 + \mathcal{G}u_2w_3 + \mathcal{D}u_3w_2 &
\mathcal{I}u_1w_2 + \mathcal{F}u_2w_1 + \mathcal{C}u_3w_3 &
\mathcal{E}u_1w_3 + \mathcal{B}u_2w_2 + \mathcal{H}u_3w_1 \\
\mathcal{H}u_1w_2 + \mathcal{E}u_2w_1 + \mathcal{B}u_3w_3 &
\mathcal{D}u_1w_3 + \mathcal{A}u_2w_2 + \mathcal{G}u_3w_1 &
\mathcal{C}u_1w_1 + \mathcal{I}u_2w_3 + \mathcal{F}u_3w_2 \\
\mathcal{F}u_1w_3 + \mathcal{C}u_2w_2 + \mathcal{I}u_3w_1 &
\mathcal{B}u_1W_1 + \mathcal{H}u_2w_3 + \mathcal{E}u_3w_2 &
\mathcal{G}u_1w_2 + \mathcal{D}u_2w_1 + \mathcal{A}u_3w_3
\end{array} \right).
\end{eqnarray}

\section{A Numerical Example}

\subsection{The SM Fermion Masses and Mixings at the GUT Scale}

The main reason for the addition of these four-point corrections
is to better match the Yukawa coupling matrices to those obtained
by running the renormalization group equations (RGEs) up to the
GUT scale. In particular, we would like to better match the lepton
masses.  It should be remembered that the parameters from the
theta function and the parameters in Table \ref{4-area} which
depend on the D-brane shift parameters are not independent. Thus,
the different elements of the Yukawa mass matrix are related, as
discussed in \cite{Chen:2007px} which strongly constrains the form
that they may take. As before, we will perform a transformation on
the SM fermion mass matrices which are obtained from the RGE
running of the experimental values for the SM fermion masses up to
the GUT scale to make the comparison with the theoretical results.
If we define $D_u$ and $D_d$ as the mass diagonal matrices of the
up- and down-type quarks respectively, the transformations are
\begin{equation}
U^u_L M_u {U^u_R}^{\dag}=D_u,\;\; U^d_L M_d {U^d_R}^{\dag}=D_d,
\;\; V_{CKM}=U^u_L {U^d_L}^{\dag},
\end{equation}
and the squared mass matrices $H_u$ and $H_d$ are
\begin{equation}
H_u = M_u M_u^{\dag}, \;\; H_d = M_d M_d^{\dag}.
\end{equation}
For simplicity, we assume that the quark mass matrices $M_u$ and
$M_d$ are Hermitian. If we take a case in which $M_d$ is very
close to the diagonal matrix, or in other words $U^d_L$ and
$U^d_R$ are very close to the unit matrix with very small
off-diagonal terms, we have
\begin{equation}
V_{CKM}\sim U^u U^{d\dag}\sim U^u,
\end{equation}
where we have transformed away the right-handed effects and make
them the same as the left-handed ones.  Then the mass matrix of
the up-type quarks turns out as
\begin{equation}
M_u \sim V_{CKM}^{\dag} D_u V_{CKM}.
\end{equation}

At the GUT scale, the CKM quark mixing matrix is given by
\cite{masses}
\begin{equation}
V_{CKM} = \left(\begin{array}{ccc}
0.9754 & 0.2205 & -0.0026i  \\
-0.2203e^{0.003^{\circ}i} & 0.9749 & 0.0318 \\
0.0075e^{-19^{\circ}i} & -0.0311e^{1.0^{\circ}i} & 0.9995
\end{array} \right),
\label{CKM-GUT}
\end{equation}
and $D_u$ and $D_d$ are
\begin{equation}
D_u = m_t \left(\begin{array}{ccc}
0.0000139 & 0 & 0  \\
0 & 0.00404 & 0 \\
0 & 0 & 1
\end{array} \right), \;\;
D_d = m_b \left(\begin{array}{ccc}
0.00141 & 0 & 0  \\
0 & 0.0280 & 0 \\
0 & 0 & 1
\end{array} \right),
\end{equation}
so that the SM fermion mass matrices (we exclude the phases in this
discussion) that we would like to match are given by
\begin{equation}
|M_u| = m_t \left(\begin{array}{ccc}
0.000266 & 0.00109 & 0.00747  \\
0.00109 & 0.00481 & 0.0310 \\
0.00747 & 0.0310 & 0.999
\end{array} \right),~~
|M_d|= D_d = m_b \left(\begin{array}{ccc}
0.00141 & 0 & 0  \\
0 & 0.0280 & 0 \\
0 & 0 & 1
\end{array} \right). \label{quarkmass}
\end{equation}
We use the relation $\frac{m_{\tau}}{m_b}=1.58$, as well as require the
eigenmasses of the leptons to be
\begin{equation}
\{ m_e, m_{\mu}, m_{\tau} \} =  m_{\tau} \{ 0.000217, 0.0458, 1\}.
\label{leptonmass}
\end{equation}

\subsection{The Numerical Results}

As mentioned above, there are six Higgs states in this model, and
only three of them can provide the SM fermion Yukawa couplings
from three-point functions on the second torus. And from Eq.
(\ref{Yukawa}) we can see explicitly only three linear
combinations of the VEVs dominate the SM fermion mass matrices. To
fit the diagonal terms in the up-type quark mass matrix, we have a
limited number of degrees of freedom to work with on the
off-diagonal mixings. Thus, it is hoped that the four-point
corrections will improve this situation. In short, the situation
for the Yukawa couplings of this model is not as ideal as the
model analyzed in~\cite{Chen:2007px} due to the highly dependent
relation between the diagonal and off-diagonal terms which results
from fewer Higgs fields. Therefore, for the best fits to the SM
fermion masses and mixings at the GUT scale, we will form the
Yukawa mass matrices such that the diagonal terms result from both
the three-point and four-point couplings, while the off-diagonal
elements arise strictly from the four-point corrections. Since we
also expect the down-type quark mass matrix to be diagonal, it is
necessary to choose reasonably small off-diagonal terms for the
mixing of the first two quark generations, so we set the K\"ahler
parameter $\frac{3J^{(2)}}{\alpha'}=30.0$.

\subsubsection{Quantum Contributions}

\begin{figure}[h]
\begin{center}
\includegraphics[width=.5\textwidth,angle=0]{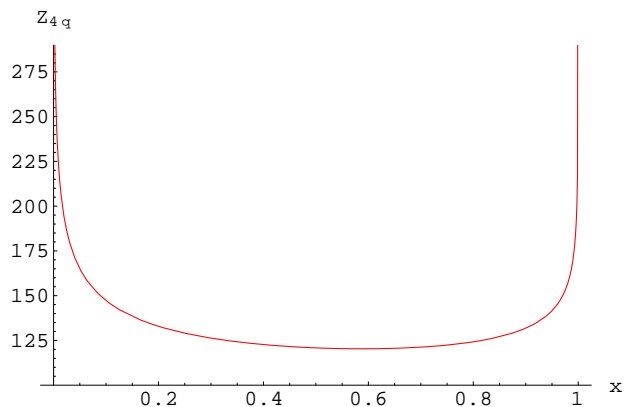}
\end{center}
\caption{$Z_{4q}$ as a function of $x$ in the range $x=(0,1)$. }
\label{Z4P}
\end{figure}

As mentioned above, the quantum contributions from the three-point
and the four-point interactions can be ignored since they are
able to be absorbed into the VEVs of the vector-like~fields:
\begin{equation}
{v'}_i^{\Phi}=Z_{3q}v^{\Phi}~;~~ u'_i w'_j=Z_{4q} u_i w_j~.
\end{equation}
However it is interesting to
study how much they may affect the VEVs.  The angles between the
branes on the second torus are
\begin{equation}
\lambda\pi=\angle ab =\angle bb'=\pi/2~;~~ \nu\pi=\angle ac
=\angle b'c =\pi/4~,
\end{equation}
so from Eqs. (\ref{Zq3}) and (\ref{Z4qu}) we have
\begin{equation}
Z_{3q}^{(2)}\sim 12.95~;~~ Z_{4q}^{(2)}(x)|_{x\sim0.5}\sim 120.8~.
\end{equation}
In this case, the plot of $Z_{4q}$ as a function of $x$ in the
range $x=(0,1)$ is shown in Figure \ref{Z4P}, where we see that $Z_{4q}$ approaches
its three-point function limits of the corresponding fields when
$x$ approaches 0 or 1 as a constant, so we confine our interest of
$Z_{4q}$ near $x=0.5$.  We find that $Z_{4q}/Z_{3q}\sim
\mathcal{O}(10)$.

\subsubsection{Yukawa Couplings from Three-Point Functions}

Let us first consider the SM fermion Yukawa couplings only from
the three-point functions. For the best fits, we set appropriate
values of $A^{\Phi}_3v^{\Phi}+D^{\Phi}_3v^{\Phi}$ to exactly match
the diagonal terms of the SM fermion mass matrices at the GUT
scale in Eqs.~(\ref{quarkmass}) and (\ref{leptonmass}). The
parameters and the VEVs in the quark and lepton mass matrices are
listed in Table \ref{3-para-p}.

\begin{table}[h]
\begin{center}
\footnotesize
\begin{tabular}{|l|c|c|c|c|c|c|@{}c@{}|@{}c@{}|@{}c@{}|} \hline

$\Phi\setminus$Parameters  & $\epsilon^{(2)}$(shift) & $A^{\Phi}$ &
$B^{\Phi}$ & $C^{\Phi}$ & $A_3^{\Phi}$ & $D_3^{\Phi}$ &
$A^{\Phi}_3v_1^{\Phi}+D^{\Phi}_3v_4^{\Phi}$ &
$A^{\Phi}_3v_2^{\Phi}+D^{\Phi}_3v_5^{\Phi}$ &
$A^{\Phi}_3v_3^{\Phi}+D^{\Phi}_3v_6^{\Phi}$  \\ \hline

$\Phi=u$ up quarks & 0 & 1 & 0.000028 & 0.000028 & 1 & 0 & 0.000266$m_t$ & 0.00481$m_t$ & 1.0$m_t$ \\
\hline

$\Phi=d$ down quarks & 0.0775 & 0.567749 & 0 & 0.002094 & 1 & 0 & 0.00248$m_b$ & 0.0493$m_b$ & 1.761$m_b$ \\
\hline

$\Phi=l$ leptons & 0 & 1 & 0.000028 & 0.000028 & 0.901 & $\sim 0$ & 0.00224$m_b$ & 0.0444$m_b$ & 1.587$m_b$\\
\hline
\end{tabular}
\caption{The matrix elements and the parameters when only trilinear
Yukawa couplings considered. Note that $v_i^d=v_i^l$. }
\label{3-para-p}
\end{center}    %T6-1-(U3)
\end{table}

The SM fermion mass matrices from the three-point function
contributions are
\begin{eqnarray}
&&|M_{3u}| \sim m_t \left(\begin{array}{@{}ccc@{}}
0.000266 & 0.000028 & 1.34\cdot10^{-7}  \\
0.000028 & 0.00481 & \sim 0 \\
1.34\cdot10^{-7} & \sim 0 & 1.0
\end{array} \right), ~
|M_{3d}| \sim m_b \left(\begin{array}{@{}ccc@{}}
0.00141 & 0.00369 & 0  \\
0 & 0.0280 & 5.19\cdot10^{-6} \\
0.000103 & 0 & 1.0
\end{array} \right), ~ \nonumber \\
&&|M_{3l}| \sim m_b \left(\begin{array}{@{}ccc@{}}
0.00224 & 0.0000444 & 1.24\cdot10^{-6}  \\
0.0000444 & 0.0444 & \sim 0 \\
1.24\cdot10^{-6} & \sim 0 & 1.568
\end{array} \right).
\end{eqnarray}

Therefore, with only three-point functions, we can obtain the
correct SM quark masses, and tau lepton mass. However, the quark
CKM mixings  are too small, and the electron and muon masses are
far from the desired values.

%%%%%%%%%%%%%%%%%%%%%%%%%%%%%%%%%%%%%%%%%%%%%%%%%%%%%%%%%%%%%%%%%%%%%%%%%%%%%%

\subsubsection{Yukawa Couplings from Three-Point and Four-Point Functions}

Let us study the SM fermion masses and mixings from both
three-point and four-point functions. To be concrete, we present
the SM fermion mass matrices from three-point and four-point
functions separately in our best fits.  First, we present the
parameters and the VEVs for three quark and lepton mass matrices
of the Yukawa couplings from the three-point functions in Table
\ref{3-para}.

\begin{table}[h]
\begin{center}
\footnotesize
\begin{tabular}{|l|c|c|c|c|c|c|@{}c@{}|@{}c@{}|@{}c@{}|} \hline

$\Phi\setminus$Parameters  & $\epsilon^{(2)}$(shift) & $A^{\Phi}$
& $B^{\Phi}$ & $C^{\Phi}$ & $A_3^{\Phi}$ & $D_3^{\Phi}$ &
$A^{\Phi}_3v_1^{\Phi}+D^{\Phi}_3v_4^{\Phi}$ &
$A^{\Phi}_3v_2^{\Phi}+D^{\Phi}_3v_5^{\Phi}$ &
$A^{\Phi}_3v_3^{\Phi}+D^{\Phi}_3v_6^{\Phi}$  \\ \hline

$\Phi=u$ up quarks & 0 & 1 & 0.000028 & 0.000028 & 1 & 0 & 0.0000250$m_t$ & 0.000262$m_t$ & 0.981$m_t$ \\
\hline

$\Phi=d$ down quarks & 0.0775 & 0.567749 & 0 & 0.002094 & 1 & 0 & 0.002820$m_b$ & 0.08784$m_b$ & 1.754$m_b$ \\
\hline

$\Phi=l$ leptons & 0 & 1 & 0.000028 & 0.000028 & 0.901 & $\sim 0$ & 0.002541$m_b$ & 0.07916$m_b$ & 1.581$m_b$\\
\hline
\end{tabular}
\caption{The matrix elements in terms of parameters. Note
$v_i^d=v_i^l$. } \label{3-para}
\end{center}    %T6-1-(U3)
\end{table}

The SM fermion mass matrices from the three-point interactions
are
\begin{eqnarray}
&&|M_{3u}^{3pt}| \sim m_t \left(\begin{array}{@{}ccc@{}}
0.0000250 & 0.0000275 & \sim 0  \\
0.0000275 & 0.000262 & \sim 0 \\
\sim 0 & \sim 0 & 0.981
\end{array} \right), ~
|M_{3d}^{3pt}| \sim m_b \left(\begin{array}{@{}ccc@{}}
0.00160 & 0.00367 & 0  \\
0 & 0.0499 & 0.00000591 \\
0.000184 & 0 & 0.996
\end{array} \right), ~ \nonumber \\
&&|M_{3l}^{3pt}| \sim m_b \left(\begin{array}{@{}ccc@{}}
0.00254 & 0.0000443 & 2.22\cdot10^{-6}  \\
0.0000443 & 0.0792 & 7.12\cdot10^{-8} \\
2.22\cdot10^{-6} & 7.12\cdot10^{-8} & 1.581
\end{array} \right)~.~\,
\end{eqnarray}

Second, for the Yukawa couplings from the four-point functions, we
present the parameters and the shifts of the D-branes in Table
\ref{4-para}, and the VEVs of three $S_L^i$ and $H^{\prime
i}_{u,d}$ in Table \ref{4-vev}.
%%%%%%%%%%%%%%%%%%%%%%%%%%%%%%%%%%%%%%%%%%%%%%%%%%%%%%%%%%%%%%%%%%%
%%%%%%%%%%%%%%%%%%%%%%%%%%%%%%%%%%%%%%%%%%%%%%%%%%%%%%%%%%%%%%%%%%%

\begin{table}[h]
\begin{center}
\footnotesize
\begin{tabular}{|l|c|c|c|c|c|c|c|c|c|c|c|} \hline

$\Phi\setminus$Parameters  & $\epsilon^{(2)}$ &
$\epsilon_{b'}^{(2)}$ & $\mathcal{A}^{\Phi}$ &
$\mathcal{B}^{\Phi}$ & $\mathcal{C}^{\Phi}$ & $\mathcal{D}^{\Phi}$
& $\mathcal{E}^{\Phi}$ & $\mathcal{F}^{\Phi}$ &
$\mathcal{G}^{\Phi}$ & $\mathcal{H}^{\Phi}$ & $\mathcal{I}^{\Phi}$
\\ \hline

$\Phi=u$ up quarks & 0 & 0.32 & 0.000064 & 0.679089 & 0.679089 &
0.000012 & 0.670140 & 0.670140 & 1.335992 & 0.000029 & 0.000029
\\ \hline

$\Phi=d$ down quarks & 0.08 & 0.03 & 0.595678 & 0 & 0.00257000 &
0.000007 & 0.028872 & 0.001674 & 0.000313 & 0.000595 & 0.073686
 \\
\hline

$\Phi=l$ leptons & 0 & 0.03 & 0.932776 & 0.000031 & 0.000031 &
0.000004 & 0.165161 & 0.165161 & 0.000171 & 0.187111 & 0.187111 \\
\hline
\end{tabular}
\caption{The matrix elements in terms of parameters. }
\label{4-para}
\end{center}
\end{table}

\begin{table}[h]
\begin{center}
\footnotesize
\begin{tabular}{|l|c|c|c|c|c|c|} \hline

$\Phi\setminus$Parameters  & $u_1 (M_s)$ & $u_2 (M_s)$ & $u_3
(M_s)$ & $w_1 (m_b)$ & $w_2 (m_b)$ & $w_3 (m_b)$
\\ \hline

$\Phi=u$ up quarks & 0.08 & 0.06 & 0.23 & 0.0148 & 0.18 & 0.003
\\ \hline

$\Phi=d$ down quarks & 0.08 & 0.06 & 0.23 & -0.004 & -0.612 & 0.03 \\
\hline

$\Phi=l$ leptons & 0.08 & 0.06 & 0.23 & -0.004 & -0.612 & 0.03 \\
\hline
\end{tabular}
\caption{The VEVs of three $S_L^i$ and $H_{u,d}^{\prime i}$. }
\label{4-vev}
\end{center}
\end{table}

The SM fermion mass matrices from the four-point interactions
are
\begin{eqnarray}
&&|M_{4u}^{4pt}| \sim m_t \left(\begin{array}{@{}ccc@{}}
0.000241 & 0.00106 & 0.00750  \\
0.00106 & 0.00455 & 0.0285 \\
0.00750 & 0.0285 & 0.0192
\end{array} \right), ~
|M_{4d}^{4pt}| \sim m_b \left(\begin{array}{@{}ccc@{}}
-0.000191 & -0.00359 & 0.0000688  \\
-0.0000361 & -0.0219 & 0.000104 \\
0.000158 & -0.00406 & 0.00409
\end{array} \right), ~ \nonumber \\
&&|M_{4l}^{4pt}| \sim m_b \left(\begin{array}{@{}ccc@{}}
-0.000299 & -0.00920 & 0.000223  \\
-0.00920 & -0.0343 & -0.0229 \\
0.00657 & -0.0229 & 0.00623
\end{array} \right).
\end{eqnarray}

Therefore, we sum over the two contributions, and obtain
\begin{eqnarray}
&&|M_{u}| \sim m_t \left(\begin{array}{@{}ccc@{}}
0.000266 & 0.00109 & 0.00750  \\
0.00109 & 0.00481 & 0.0285 \\
0.00750 & 0.0285 & 1.0
\end{array} \right), ~
|M_{d}| \sim m_b \left(\begin{array}{@{}ccc@{}}
0.00141 & 0.0000828 & 0.0000688  \\
-0.0000361 & 0.028 & -0.0000979 \\
0.0000258 & -0.00406 & 1.0
\end{array} \right), ~ \nonumber \\
&&|M_{l}| \sim m_b \left(\begin{array}{@{}ccc@{}}
0.00224 & -0.00915 & 0.000223  \\
-0.00915 & -0.0449 & -0.0229 \\
0.000223 & -0.0229 & 1.587
\end{array} \right),
\end{eqnarray}
and the mass eigenvalues of the leptons are
\begin{equation}
\{ m_e, m_{\mu}, m_{\tau} \} = m_{b} \{ 0.000348, 0.0464, 1.588\}=
m_{\tau} \{ 0.000219, 0.0292, 1\}.
\end{equation}

Comparing with the SM fermion masses and mixings in Eqs.
(\ref{quarkmass}) and (\ref{leptonmass}), we find that the
off-diagonal terms of the up-type quark mass matrix are basically
from the four-point contributions, and it is quite difficult to fit
exactly the same off-diagonal terms due to the four-point function
matrix structure. For the down-type quark matrix, we still cannot
eliminate the off-diagonal terms. However, we can suppress them such
that they are in an acceptable range.  In addition, the quark CKM
mixing matrix is
\begin{eqnarray}
&&V_{CKM} \simeq \left(\begin{array}{@{}ccc@{}}
0.977 & 0.212 & 0.00109  \\
0.212 & 0.977 & 0.0298 \\
0.00738 & 0.0289 & 0.9996
\end{array} \right),
\end{eqnarray}
and then there are some deviations for the quark mixing terms in
Eq. (\ref{CKM-GUT}), but they are in an acceptable range. Finally,
we are able to decrease the electron mass eigenvalue to the
correct value by the off-diagonal terms from the four-point
contributions, but the muon mass eigenvalue is still about 36\%
smaller than the desired value.

\section{Conclusion}

We have discussed corrections to the SM fermion Yukawa couplings
in intersecting D6-brane models due to four-point interactions and
presented a working example, and demonstrated that these
corrections can improve the best fits for the SM fermion masses
and mixings. In a concrete model, we first calculated the SM
fermion masses and mixings from three-point functions. Considering
only these contributions, we can obtain the correct quark masses
and tau lepton mass, but the CKM quark mixings are not large
enough, and the electron and muon masses are far from the desired
values. After including the corrections to the SM fermion Yukawa
couplings from four-point functions, we can obtain the correct
quark masses and CKM mixings, and the correct electron and tau
lepton mass scales. However, the muon mass is still around 36\%
smaller than the desired value.

In this work, we have not considered the moduli stabilization
problem, and have essentially treated the moduli VEVs as free
parameters. As is obvious, there exists fine-tuning in our
discussion of the D-brane positions and the Higgs VEVs. The
four-point interactions have nine dependent worldsheet instanton
parameters as well as six additional string-scale Higgs particles
and their VEVs.  To stabilize these undetermined variables, one
may consider a $\mathbf{\Z_2\times \Z_2'}$ orientifold model with
discrete torsion where the D-branes wrap rigid cycles, thus
stabilizing the open-string moduli.  In this case, one would also
expect corrections to the Yukawa couplings from D-brane
instantons. Indeed, E2-branes required for this construction must
also wrap rigid cycles, so there is additional motivation to
consider this background.  We plan to pursue these possibilities
in our future research.

\section*{Acknowledgements}

This research was supported in part by the Mitchell-Heep Chair in
High Energy Physics (CMC), by the Cambridge-Mitchell Collaboration
in Theoretical Cosmology (TL), and by the DOE grant
DE-FG03-95-Er-40917 (DVN).

\end{document}